\documentclass[prb,twocolumn,showpacs,amsmath,floatfix]{revtex4}
\usepackage{graphicx}
\usepackage{bm}

\begin{document}

\title{Fluctuation-enhanced frequency mixing in a nonlinear 
micromechanical oscillator}

\author{H. B. Chan}
\email{hochan@phys.ufl.edu}
\author{C. Stambaugh}

\affiliation{Department of Physics, University of Florida, Gainesville, FL 32611}

%
\begin{abstract}
We study noise-enhanced frequency mixing in an underdamped micromechanical torsional oscillator. The oscillator is electrostatically driven into bistability by a strong, periodic voltage at frequency  $\omega_d$. A second, weak ac voltage is applied at a frequency  $\omega$ close to $\omega_d$. Due to nonlinearity in the system, vibrations occur at both $\omega$  and $2\omega_d-\omega$. White noise is injected into the excitation, allowing the system to occasionally overcome the activation barrier and switch between the two states.  At the primary drive frequency where the occupations of the two states are approximately equal, we observe noise-induced enhancement of the oscillation amplitudes at both $\omega$ and the down-converted frequency $2\omega_d-\omega$, in agreement with theoretical predictions. Such enhancement occurs as a result of the noise-induced interstate transitions becoming synchronous with the beating between the two driving frequencies.  
\end{abstract}

\maketitle


One of the important uses of nonlinear systems is to mix frequencies. Well-known examples are higher harmonic generation and four-wave mixing in nonlinear optics. Frequency mixing is also of great importance in high sensitivity detection and signal processing. The conversion efficiency depends on a number of factors, such as the inherent nonlinearity of the system. Strong driving fields are often necessary to produce the converted signal. Certain nonlinear systems, such as Duffing oscillators, develop bistability when driven periodically beyond a critical oscillation amplitude \cite{1}. In the presence of fluctuations these systems can occasionally overcome the activation barrier in phase space and switch between the dynamical states \cite{2}. Noise activated switching has been studied in nonequilibrium systems \cite{3} including parametrically driven electrons in a Penning trap \cite{4}, radio frequency driven Josephson junctions \cite{5} and micro- and nano-mechanical oscillators \cite{6,7,8,9}. Theoretical predictions suggest that when parameters are chosen such that the occupations of the two states are equal, the interplay between noise and nonlinearity leads to a range of critical kinetic phenomena \cite{2,10}. In particular, noise could play a constructive role in enhancing the mixing of oscillations at different frequencies.

\begin{figure}
\includegraphics[angle=0]{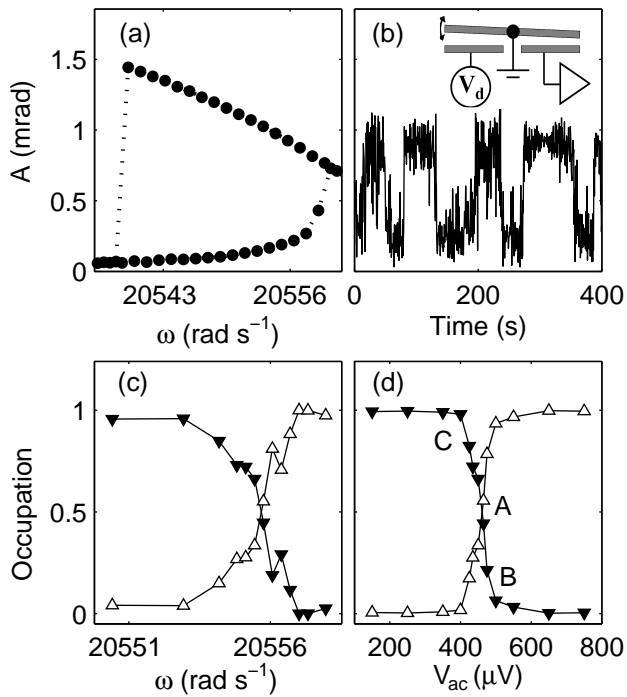} 
\caption{\label{fig:1}(a) Oscillation amplitude A vs. driving frequency. Bistability occurs between 20539.25 $\mathrm{rad\; s}^{-1}$ and 20559.97 $\mathrm{rad\; s}^{-1}$. (b) At driving frequency of 20555.75 $\mathrm{rad\; s}^{-1}$, the oscillation amplitude switches between two distinct values when noise is injected into the system. Inset: cross-sectional schematic of the micromechanical torsional oscillator.  (c) Occupations of the high amplitude state (upright hollow triangles) and the low-amplitude state (inverted solid triangles) as a function of driving frequency at fixed amplitude. (d) Occupations as a function
of the driving amplitude $\mathrm{V}_\mathrm{ac}$ at fixed driving frequency.}
\end{figure}  

In an earlier experiment \cite{9}, we drove a nonlinear micromechanical torsional oscillator into bistability and showed that in the presence of noise, there exists a narrow range of driving frequencies where the occupations of the two oscillation states are comparable. A narrow and sharp peak at the driving frequency $\omega_d$ was observed at this ``kinetic phase transition'' due to noise induced switching between the two states. In this paper, we demonstrate that when a second, weak periodic trial excitation with frequency $\omega \sim \omega_d$ is applied in addition to the strong periodic drive, fluctuation-induced interstate transitions become synchronous with the beating between the two driving frequencies at certain noise intensities, resulting in enhanced response at both the frequency of the weak excitation and the down-converted frequency. Responses at both frequencies initially increase with noise and subsequently decrease as the noise intensity becomes very strong. The occurrence of such ``high frequency stochastic resonance'' is in agreement with theoretical predictions \cite{2,10}.

\begin{figure}
\includegraphics[angle=0]{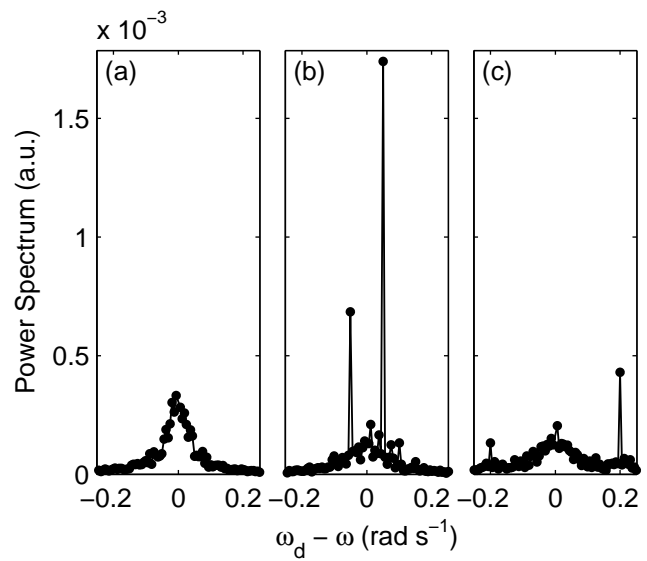} 
\caption{\label{fig:2}(a) At the driving frequency where the occupations of the two states are equal, a sharp peak develops in the power spectral density of fluctuations. (b) Addition of a secondary, weak excitation at  $\omega_1 = \omega_d + 0.05\; \mathrm{rad\; s}^{-1}$ leads to spectral peaks at $\omega_1$  and $2\,\omega_d-\omega_1$. (c) The frequency of the secondary excitation is changed to  $\omega_2 = \omega_d + 0.20\; \mathrm{rad\; s}^{-1}$.}
\end{figure}

Our micromechanical oscillator consists of a movable silicon plate supported by two torsional springs. As shown in the inset of Fig. \ref{fig:1}b, a driving voltage $\mathrm{V}_\mathrm{d}$ is applied to an electrode underneath the top plate. Torsional oscillations are excited by the periodic component of $\mathrm{V}_\mathrm{d}$ with amplitude $\mathrm{V}_\mathrm{ac} \simeq 460 \mu \mathrm{V}$ and frequency $\omega_d$. A second electrode is used to detect the oscillations capacitively. The oscillator has a resonance frequency $\omega_R$ of 20564.51 $\mathrm{rad\;s}^{-1}$ and a quality factor Q of $\sim 8100$. Details about the oscillator can be found in Refs. [\onlinecite{8,11}]. As a result of the strong distance dependence of the electrostatic attraction, the system behaves like a Duffing oscillator with cubic nonlinearity. Under sufficiently strong periodic excitation, the oscillator possesses two oscillation states with distinct amplitude and phase over a range of frequencies (Fig. \ref{fig:1}a).  The presence of noise enables the system to occasionally overcome the activation barrier and switch between these two states. Figure \ref{fig:1}b shows that the oscillation amplitude jumps between two values as a function of time when noise is injected into the excitation voltage. The occupations of the two states depend on the driving frequency. As shown in Fig. \ref{fig:1}c, the oscillator resides predominantly in the high- or low-amplitude state at the high and low frequency sides of the hysteresis loop respectively. The occupations of the two states are comparable only at a small range of frequencies. As we showed in an earlier experiment \cite{9}, noise induced interstate transitions at this ``kinetic phase transition'' lead to a narrow peak in the fluctuation spectrum that is centered at the driving frequency (Fig. \ref{fig:2}a). The width of this peak is inversely proportional to the residence time of the states and its height drops exponentially as the excitation frequency is tuned away from the value at which the occupation of the two states are comparable. If the excitation amplitude is varied with the driving frequency kept constant, the occupations of the two states exhibit a similar behavior: the occupation of the low amplitude state is high at small driving amplitude and vice versa for large driving amplitude. The occupations of the two states become comparable at some intermediate amplitude (point A in Fig. \ref{fig:1}d).

We investigate the response of the oscillator when a weak periodic drive ($\sim 120 \mu \mathrm{V}$) at frequency $\omega$ close to $\omega_d$ is applied on top of the strong periodic drive. Since the nonlinearity of the oscillator mixes the primary and secondary driving frequencies, the weak drive induces oscillations not only at its own frequency $\omega$, but also at other frequencies. The response to the weak drive is the strongest at the frequencies $\omega$ and $2\,\omega_d - \omega$. We found that at the ``kinetic phase transition'' where the occupations of the two oscillation states are equal, the presence of an optimal amount of noise enhances the mixing, resulting in amplified response at both $\omega$ and $2\,\omega_d - \omega$, in agreement with predictions from theoretical analysis \cite{10}. Figures \ref{fig:2}b and \ref{fig:2}c show the spectral response of the oscillator when weak periodic drives at two different frequencies ($\omega_1 = \omega_d + 0.05 \;\mathrm{rad\; s}^{-1}$ and $\omega_2 = \omega_d + 0.20 \;\mathrm{rad\; s}^{-1}$ respectively) were applied in the presence of noise. Spectral peaks at both the weak driving frequencies $\omega_{1,2}$ and the down-converted frequencies $2 \omega_d - \omega_{1,2}$ can be clearly identified. The peaks are about a factor of 6 stronger compared to the case when noise is not applied to the excitation. 
In both cases, the frequency difference $\omega_{1,2} - \omega_d$ is much smaller than the relaxation rate of the oscillator ($\omega_{R}/2Q \sim 1.3\, \mathrm{s}^{-1}$). As shown in Fig. \ref{fig:2}c, both spectral peaks diminish when the frequency   of the weak drive moves further apart from the frequency $\omega_d$ of the primary drive. 

\begin{figure}
\includegraphics[angle=0]{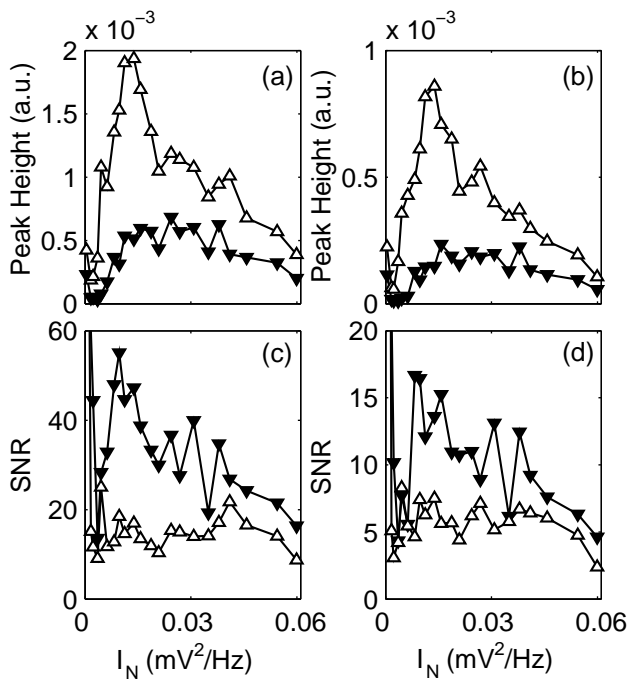} 
\caption{\label{fig:3}(a) The height of the spectral peak at  $\omega$ as a function of the noise intensity for  $\omega_1 = \omega_d + 0.05 \;\mathrm{rad\; s}^{-1}$ (upright hollow triangles) and  $\omega_2 = \omega_d + 0.20 \;\mathrm{rad\; s}^{-1}$ (inverted solid triangles). (b) The spectral peak at the down-converted frequency of $2\,\omega_d-\omega_{1,2}$  (c) The signal to noise ratios at $\omega_{1,2}$ as a function of the noise intensity. (d) The signal to noise ratio at $2\,\omega_d-\omega_{1,2}$.}
\end{figure}

Next, we examine the dependence of the spectral peaks on the noise intensity. Figure \ref{fig:3}a shows that the spectral peaks at $\omega$ first increase with noise intensity, attaining a maximum at a noise power of $\sim 0.012 \mathrm{mV}^2/\mathrm{Hz}$ and subsequently decrease. Figure \ref{fig:3}b demonstrates a similar behavior for the spectral peak at the down-converted frequency $2\,\omega_d - \omega$ where an optimal amount of noise enhances the frequency mixing. 
We emphasize that there is no periodic excitation at $2\,\omega_d - \omega$ and the spectral response in Fig. \ref{fig:3}b arises from the nonlinear mixing of the primary and secondary frequencies. Figures \ref{fig:3}c and \ref{fig:3}d plot the signal to noise ratio, defined as the spectral responses at $\omega$ and $2\,\omega_d - \omega$ divided by the power spectrum in the absence of the weak periodic excitation (Fig. \ref{fig:2}a) at the corresponding frequencies. At both $\omega$ and $2\,\omega_d - \omega$, the signal to noise ratios achieve maxima at some intermediate noise intensity.

\begin{figure}
\includegraphics[angle=0]{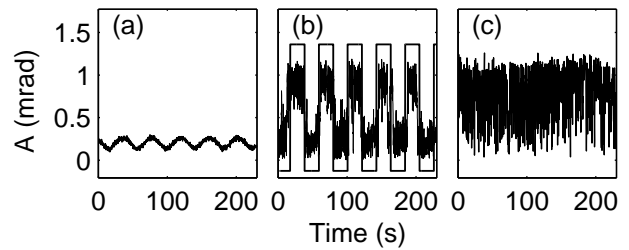} 
\caption{\label{fig:4}(a) Oscillation amplitude as a function of time in the absence of noise. (b) Under optimal noise, the oscillation amplitude switches between two stable values as a function of time, in sync with the beating frequency (square wave represented by the solid line) (c) With large intensity of noise, interstate switching loses synchronization with the beating frequency. }
\end{figure}

The enhancement of the spectral responses at $\omega$ and $2\,\omega_d - \omega$ was predicted to occur when noise-induced interstate transitions become synchronous with the beating between the primary and secondary driving frequencies \cite{10}. 
For an intuitive understanding, we consider the beating between the primary and secondary drives that leads to modulation of the amplitude of the strong field at frequency $\left | \omega - \omega_d \right|$. Since the relaxation rate of the oscillator ($\omega_{R}/2Q \sim 1.3\, \mathrm{s}^{-1}$) is much larger than $\left | \omega - \omega_d \right|$, the oscillation amplitude at the primary frequency varies adiabatically with the driving amplitude and is therefore modulated at the beating frequency (Fig. \ref{fig:4}a). In the absence of noise, the oscillator remains in the low-amplitude state provided that the modulation envelope is small. When sufficient noise is introduced into the system, the oscillator switches between the two oscillation states (Fig. \ref{fig:4}b). Slow amplitude modulations of the primary excitation change the probability of fluctuational transitions and hence the occupations of the two states. 
As illustrated in Fig. \ref{fig:1}d, the occupation probability of the low-amplitude state, for example, decreases as the excitation amplitude increases (point B) and vice versa (point C). The occupation probabilities hence vary periodically as a function of time. At the optimal noise intensity, inter-state transitions occur approximately once every half-cycle of the beating (Fig. \ref{fig:4}b). Therefore, the amplitude modulation of the response becomes significantly larger compared to Fig. \ref{fig:4}a, resulting in spectral components at both the secondary frequency $\omega$ and the down-converted frequency $2\,\omega_d - \omega$ that are enhanced compared to the case when noise is absent. 
Upon further increase of the noise intensity, the transition rate between the two states exceeds the beating frequency and the switching events are no longer synchronous with the modulation in driving amplitude (Fig. \ref{fig:4}c). As a result, the spectral response at $\omega$ and $2\,\omega_d - \omega$ decreases.

\begin{figure}
\includegraphics[angle=0]{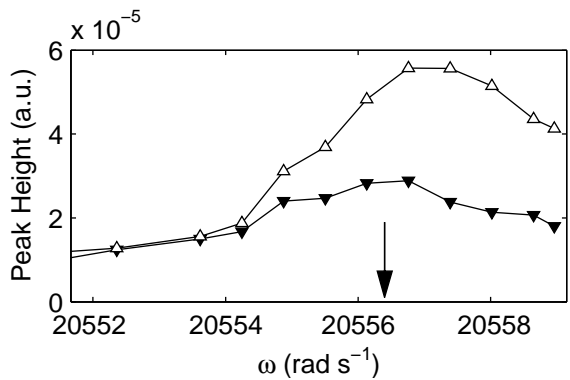} 
\caption{\label{fig:5}Dependence of the spectral power at  $\omega_1$ (upright hollow triangles) and $2\,\omega_d-\omega_1$ (inverted solid triangles) on the driving frequency  $\omega_d$.  The frequency difference between the primary and secondary drives ($\omega_1-\omega_d$) is maintained constant. The arrow indicates the frequency at which the occupations of the two
states are equal. This frequency is slightly shifted from Fig. \ref{fig:1} because the
device has been warmed up to room temperature and re-cooled down.}
\end{figure}

The noise enhancement of spectral responses at $\omega$ and $2\,\omega_d - \omega$ occurs only in the vicinity of the ``kinetic phase transition'' where the occupations of the two states are comparable. Figure \ref{fig:5} shows the spectral responses at $\omega$ and at $2\,\omega_d - \omega$ as a function of $\omega_d$, where the frequency difference between the primary and secondary drives ($\omega - \omega_d$) is maintained constant. 
The spectral responses at both frequencies attains maximum when the occupations of the two states are nearly equal. As the driving frequency is moved away from the  ``kinetic phase transition'', the occupation of one of the states become much larger than the other and interstate transitions becomes significantly less frequent. As a result, the enhancement of spectral responses becomes weaker.

Such ``high frequency stochastic resonance'' was predicted to occur when a weak, periodic drive is applied to systems where bistability arises as a result of a primary, strong periodic modulation \cite{10}. For ordinary systems with bistable potential, stochastic resonance \cite{12,13} takes place when the rate of noise-induced transitions becomes comparable to the frequency of the periodic drive. Our oscillator, in contrast, is monostable under weak periodic driving and develops bistability only when it is driven strongly. Both the primary and secondary periodic drives are applied at frequencies much higher that the rate of noise induced transitions. Another important difference from conventional stochastic resonance is that noise-enhanced response also takes place at $2\,\omega_d - \omega$ at which no periodic excitation is applied. Such efficient down-conversion of excitation frequency occurs due to synchronization of the switches in the oscillation amplitude with the beating frequency.

Critical kinetic phenomena such as the emergence of the narrow peak at the driving frequency \cite{2,9,10} and noise enhanced frequency mixing reported here occur as a result of the interplay between noise and nonlinearity. Other bistable systems far from equilibrium, including rf-driven Josephson junctions \cite{5}, nanomagnets driven by polarized current \cite{14} and double barrier resonant tunneling structures \cite{15}, are expected to exhibit similar behavior in the parameter range where the two states have comparable occupations. The investigation of these critical kinetic phenomena could also open up new opportunities in tunable narrow band filtering and detection using micromechanical oscillators.

We thank M. I. Dykman and D. Ryvkine for useful discussions.



\end{document}